\documentclass[fp, twocolumn]{jpsj3}
\usepackage[dvipdfmx]{graphicx}
\usepackage{amsmath}
\usepackage{amsfonts}
\usepackage{dcolumn}
\usepackage{bm}
\usepackage{physics}
\usepackage{color}

\title{
    Enhanced \(2\pi\)-periodic Aharonov-Bohm Effect as a Signature of Majorana Zero Modes Probed by Nonlocal Measurements
}

\author{Masayuki Sugeta, Takeshi Mizushima, and Satoshi Fujimoto}
\inst{Department of Materials Engineering Science, Osaka University, Toyonaka, Osaka 560-8531, Japan}
\abst{
    We propose the \(2\pi\)-periodic Aharonov-Bohm (AB) effect as a nonlocal probe of Majorana zero modes (MZMs) without the restriction of fermion parity. We demonstrate the enhancement of the AB effect, where the topological protection of MZMs yields amplified and robust Andreev reflection mediated by MZMs at multiple superconductor-normal metal junctions. We investigate the influence of trivial bound states and show that a nonlocal index enables a more explicit distinction between the trivial and topological bound states than local probes.
}

\begin{document}
\maketitle

\section{Introduction}
Owing to exotic properties such as topological protection and non-Abelian braidings, Majorana zero modes (MZMs) have been one of central topics in condensed matter physics in the last decade. The potential application for fault-tolerant quantum computation~\cite{2008_Nayak, 2020_Aguado} motivates the pursuit of MZMs. One of the most promising platforms to host MZMs is a class of materials called topological superconductors. Several physical systems have been proposed to realize the topological bound states, including Rashba spin-orbit coupled nanowires with Zeeman energy in the proximity of \textit{s}-wave superconductors~\cite{2008_Fujimoto, 2009_Sato, 2010_Sau, 2010_Alicea, 2010_Sato, 2010_Lutchyn, 2010_Oreg, 2012_Alicea, 2016_Sato, 2017_Aguado, 2017_Kiczek, 2020_Prada}. This system is particularly favorable to experimentalists thanks to the emergence of the exotic property from the combination of conventional ingredients.

Despite the substantial effort to prove the existence of MZMs, the consensus is yet to be established. The main focus of previous experimental studies has been local conductance measurements~\cite{2012_Mourik, 2012_Das, 2012_Deng, 2013_Finck, 2016_Deng, 2017_Chen, 2018_Gul, 2021_Zhang}. The topologically protected Andreev reflection mediated by MZMs causes a quantized zero-bias conductance peak (ZBCP)~\cite{2012_Alicea, 2016_Sato}, which is one of the most well-known signatures of MZMs. For example, the experimentally observed near-quantized ZBCPs~\cite{2017_Nichele} were interpreted as evidence of MZMs. However, theoretical studies have pointed out that the trivial bound states called partially-separated Andreev bound states (ps-ABSs) can mimic the signature~\cite{2012_Kells, 2012_Prada, 2015_Cayao, 2016_San-Jose, 2017_Ptok, 2018_Fleckenstein, 2018_Moore_1, 2018_Moore_2, 2018_Reeg, 2018_Penaranda, 2019_Dvir, 2019_Vuik, 2019_Avila, 2020_Dmytruk, 2020_Junger, 2020_Prada, 2020_Razmadze, 2021_Valentini, 2021_Cayao_2, 2021_Yu}. The indistinguishability of the signatures emerging from MZMs and ps-ABSs questions the validity of the previous studies based on local conductance measurements, motivating exploration for alternative experimental schemes to distinguish the trivial and topological bound states~\cite{2018_Liu, 2019_Ricco, 2019_Yavilberg, 2019_Awoga, 2020_Schulenborg, 2020_Zhang, 2021_Pan, 2021_Liu, 2021_Ricco, 2021_Thamm, 2022_Chen}.

The methods to probe the nonlocality peculiar to MZMs, among nonlocal transport measurements~\cite{2017_Soori, 2019_Nehra, 2019_Valkov, 2019_Soori, 2020_Soori, 2020_Danon, 2021_Cayao_1}, include the ``teleportation'' interferometry~\cite{2007_Semenoff, 2007_Bolech, 2008_Tewari, 2008_Nilsson, 2010_Fu}. The process is the phase-coherent single electron transmission through MZMs, which reflects the nonlocal nature as the independence of the distance between MZMs~\cite{2010_Fu}. In this process, the phase of the transmission amplitude is proportional to the occupation number of the complex fermion composed of the MZMs. The interference measurement not only provides a smoking-gun signature of MZMs but also serves as readouts of topological qubits~\cite{2016_Landau, 2016_Plugge}.

While the interferometry scheme attracts considerable attention from both theoretical and experimental points of view~\cite{2011_Zazunov, 2012_Hutzen, 2014_Rainis, 2016_Vijay, 2018_Hell, 2019_Liu, 2020_Whiticar}, it requires parity conservation~\cite{2010_Fu}. The condition demands the fabrication of mesoscopic systems with floating superconductors, which may lead to experimental intricacy. Therefore, it is desirable to develop a new approach for detecting the nonlocality of MZMs without the condition of parity restrictions.

In this paper, we propose the \(2\pi\)-periodic Aharonov-Bohm (AB) effect as a nonlocal probe of MZMs without the requirement of parity restrictions. We introduce two indices that characterize the AB effect and discuss their capability for distinguishing MZMs from ps-ABSs. We discuss how the robustness of the indices demonstrates the difference between the topological and trivial bound states. We show that the effect can exhibit a more conspicuous contrast between topological and trivial bound states than local conductance measurements.

\section{Basic Idea of \(2\pi\)-periodic AB Effect as a Probe of Majorana Zero Modes}
\label{section_2piAB}
\begin{figure*}[t]
    \includegraphics[width=\textwidth]{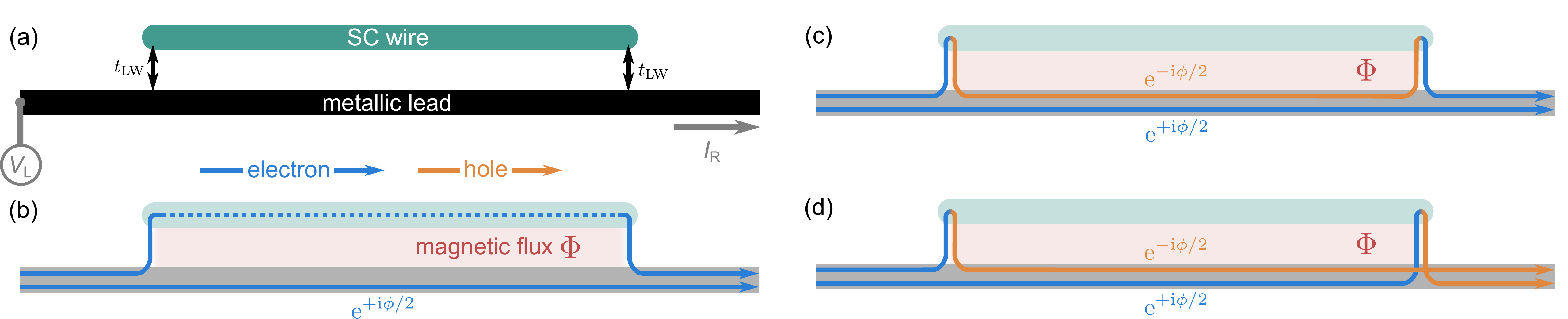}
    \caption{
        (Color online) 
        (a) Setup for this research. The one-dimensional topological superconductor is placed parallel to the metallic lead. The lead and the superconducting wire are connected at both ends of the wire, forming the closed-loop system with SN junctions.
        (b) Interference process with a floating superconductor. The teleportation only occurs in the topological phase. (c) Interference process with a grounded superconductor. Electrons are doubly Andreev-reflected at the left and right junctions and interfere with the original electrons. (d) Another process with a grounded superconductor. Electrons are Andreev-reflected at the left or right junction. The holes created at each junction interfere with each other.        
        }
    \label{fig_setup_process}
\end{figure*}

We first briefly review the AB effect in one-dimensional loop systems with a superconducting segment, followed by an explanation of the basic idea of this paper: \(2\pi\)-periodic AB effect as a probe of MZMs.
Here we consider the AB effect in the setup expressed in Fig.~\ref{fig_setup_process}~(a) by introducing Peierls phase factors to the hopping of the metallic lead in the tight-binding model. 
Under the condition of parity restriction, MZMs at the ends of the wire mediate teleportation in the topological phase. The interference between the two paths of electron conduction, namely the metallic lead and the superconducting wire, yields the \(4\pi\)-periodic AB effect~\cite{2010_Fu}. 
In the trivial phase, the absence of transmission channels inside the energy gap eliminates the effect (see Fig.~\ref{fig_setup_process}~(b)).

This paper is devoted to the case without parity restriction, where the AB effect is \(2\pi\)-periodic~\cite{2010_Benjamin, 2022_Valkov, 2022_Aksenov}. The halved periodicity is attributed to the two processes with Andreev reflections at both ends of the superconducting wire (see Fig.~\ref{fig_setup_process}).
The Andreev reflections at both superconductor-normal metal (SN) junctions cause electrons and holes to gain the opposite-signed phase, resulting in the \(2\pi\) periodicity. The halved periodicity is universal whether the system is in the topological or trivial phase. However, the amplitude is more conspicuous by the orders of magnitude in the topological phase because the topological bound states mediate the perfect Andreev reflection~\cite{2011_Akhmerov, 2011_Fulga}. Due to the spatial separation and the topological protection of MZMs, the processes of multiple Andreev reflections via MZMs are more robust than those via independent unprotected states at SN junctions. Although the effect solely relies on local Andreev reflections, it can provide a more striking difference between topological and trivial bound states because the processes depend on the conditions of multiple SN junctions. That is, in the case of the topological superconducting state, there are topologically protected MZMs at both two ends, and hence, the sequential Andreev reflections mediated via these two MZMs can maintain the interference effect robustly. On the other hand, in the trivial case, for preserving the interference effect due to the sequential scatterings at two junctions, one needs to tune system parameters precisely to make the energy levels of two ps-ABSs at two ends nearly equal. In the following sections, we confirm these predictions by performing precise model calculations.

\section{Model}
We model a topological superconductor as a Rashba spin-orbit coupled nanowire with Zeeman energy and \textit{s}-wave superconducting proximity effect. The Hamiltonian is expressed as~\cite{2010_Sato}
\begin{equation}
	\mathcal{H} = \mathcal{H}_\mathrm{wire} + \mathcal{H}_\mathrm{lead} + \mathcal{H}_\mathrm{hop\text{-}LW},
    \label{eq_Hamiltonian_type}
\end{equation}
\begin{eqnarray}
	\mathcal{H}_\mathrm{wire} 
	&=& \sum_{j, \sigma} \left[-t_\mathrm{wire} c_{j + 1, \sigma} ^\dagger c_{j, \sigma} + \mathrm{h.c.}\right] \nonumber \\
    && - \mu_\mathrm{wire} \sum_{j, \sigma} c_{j, \sigma}^\dagger c_{j, \sigma} + h \sum_j \left[c_{j, \uparrow} ^\dagger c_{j, \uparrow} - c_{j, \downarrow}^\dagger c_{j, \downarrow}\right] \nonumber \\
	&& + \sum_j \left[-\lambda c_{j - 1, \downarrow}^\dagger c_{j, \uparrow} + \lambda c_{j + 1, \downarrow}^\dagger c_{j, \uparrow} + \mathrm{h.c.}\right] \nonumber \\
	&& + \sum_j \left[\Delta c_{j, \uparrow}^\dagger c_{j, \downarrow}^\dagger + \mathrm{h.c.}\right],
    \label{eq_Hamiltonian_wire}
\end{eqnarray}
\begin{eqnarray}
	\mathcal{H}_\mathrm{lead} 
	&=& \sum_{j, \sigma} \left[-t_\mathrm{lead} \psi_{j + 1, \sigma} ^\dagger \psi_{j, \sigma} + \mathrm{h.c.}\right] \nonumber \\
    && - \mu_\mathrm{lead} \sum_{j, \sigma} \psi_{j, \sigma}^\dagger \psi_{j, \sigma}
    \label{eq_Hamiltonian_lead},
\end{eqnarray}
\begin{equation}
	\mathcal{H}_\mathrm{hop\text{-}LW} = \sum_\sigma \left[- t_\mathrm{LW} c_{1, \sigma}^\dagger \psi_{1, \sigma} - t_\mathrm{LW} c_{N, \sigma}^\dagger \psi_{N, \sigma} + \mathrm{h.c.}\right]
    \label{eq_Hamiltonian_hop}.
\end{equation}
Throughout this paper, we set parameters \(\lambda = 0.3 t_\mathrm{wire}\), \(\Delta = 0.1 t_\mathrm{wire}\), \(t_\mathrm{lead} = t_\mathrm{wire}\), \(\mu_\mathrm{lead} = 0\). The length of the nanowire is \(L_\mathrm{wire} = 500a\) unless stated otherwise, where \(a\) is the lattice constant of the tight-binding model.
The nonlocal conductance \(G = d I_\mathrm{R} / d V_\mathrm{L}\) (see Fig.~\ref{fig_setup_process}~(a)) is calculated as,
\begin{equation}
	G = \frac{e^2}{h} \int_{-\infty}^\infty dE \left(- \pdv{f(E - eV)}{E}\right) \left[T_\mathrm{ee}(E) - T_\mathrm{eh}(E)\right].
    \label{eq_Landauer-Buttiker}
\end{equation}
Here \(T_\mathrm{ee}\) (\(T_\mathrm{eh}\)) is the transmission rate from electron sectors of the left lead to electron (hole) sectors of the right lead. The transmission rates are calculated with the help of Kwant, which is the Python package for the scattering problem in tight-binding models~\cite{2014_Groth}.

\section{Results}
\subsection{Pristine Nanowire}
\begin{figure}[t]
    \includegraphics[width=85mm]{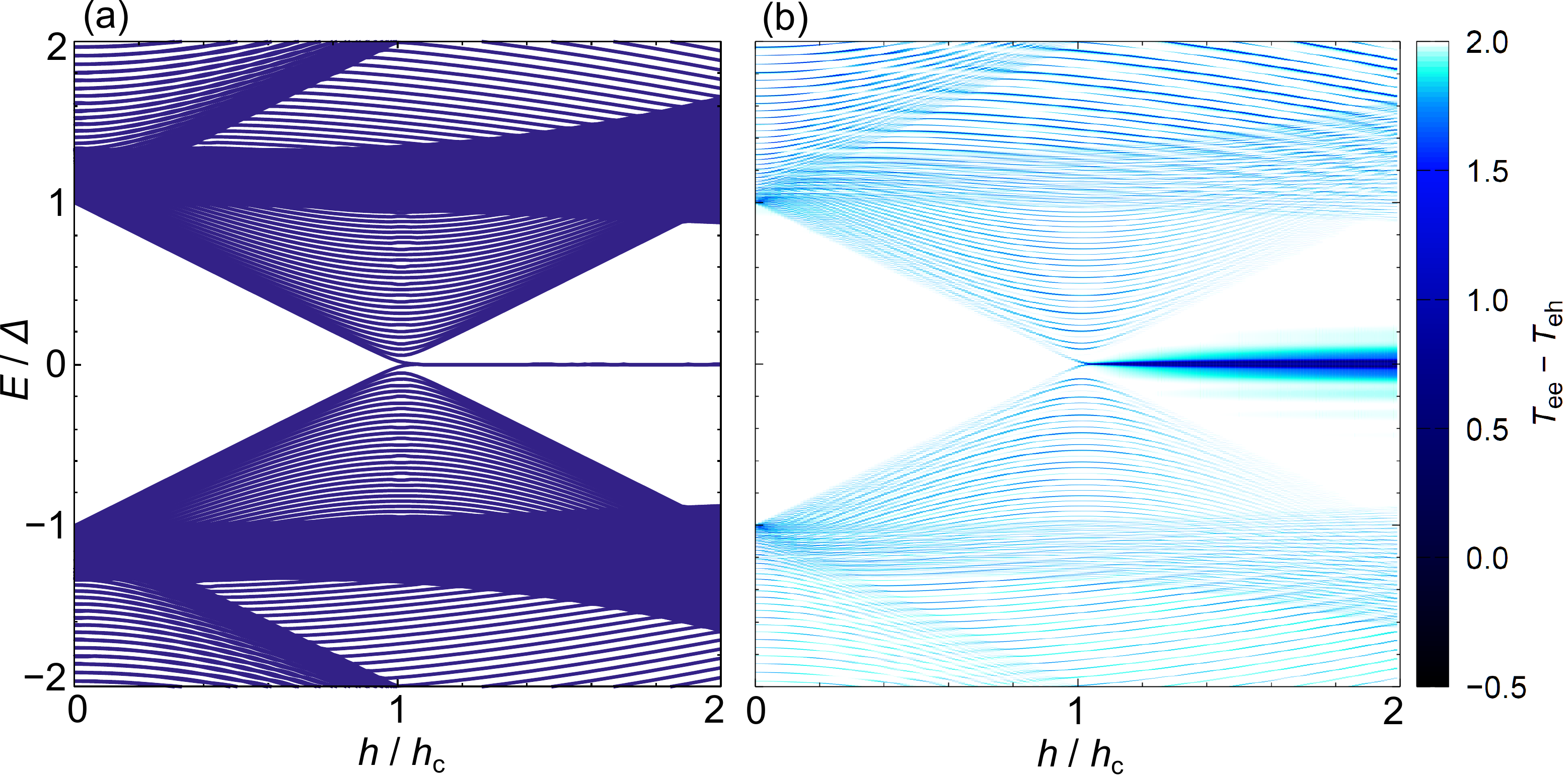}
    \caption{
        (Color online) 
        (a) Magnetic field dependence of the energy eigenvalues of the isolated superconductor. (b) Magnetic field dependence of the transmission rate (\(T_\mathrm{ee} - T_\mathrm{eh}\)), or the conductance in the unit of \(e^2 / h\) at zero temperature. Parameters are set to be \(t_\mathrm{LW} = 0.25 t_\mathrm{lead}\), \(\mu_\mathrm{wire} = -2 t_\mathrm{wire}\).
        }
    \label{fig_h-E-G_no-psABSs}
\end{figure}

Figure~\ref{fig_h-E-G_no-psABSs}~(b) shows the magnetic field dependence of the conductance of the SN junction system. 
The system undergoes topological phase transition by changing magnetic field \(h\), with the critical value of \(h_\mathrm{c}\) (see Fig.\ref{fig_h-E-G_no-psABSs}~(a)). The trivial system (\(h < h_\mathrm{c}\)) yields a nonzero and constant conductance inside the energy gap.
The conductance reflects the gap-closing behavior with increasing \(h\). After the phase transition and the emergence of MZMs, the conductance shows the valley structure at zero energy. The emergence of the structure can be interpreted as a signature of the topological phase.

\begin{figure}[t]
    \includegraphics[width=85mm]{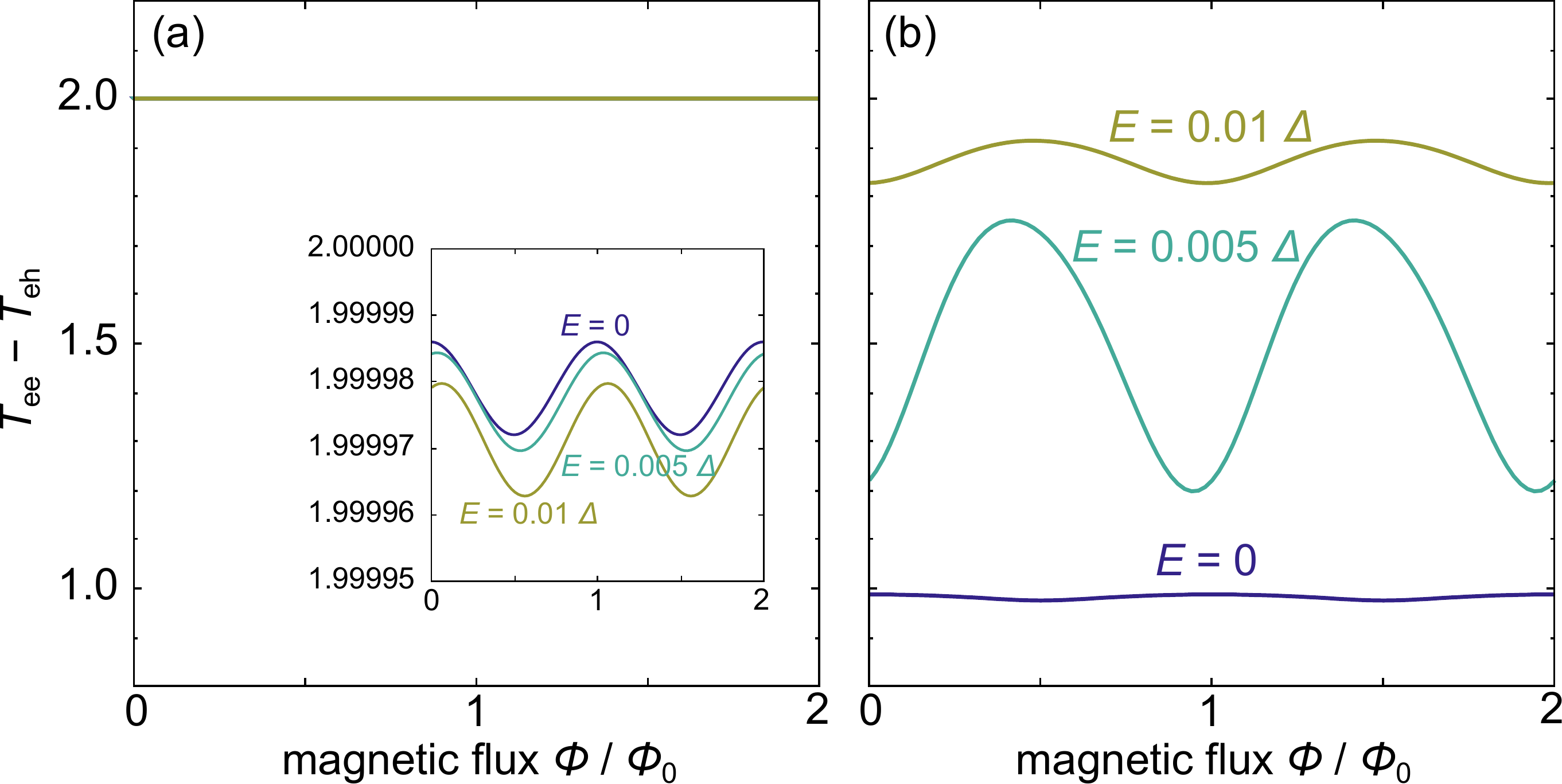}
    \caption{
        (Color online) 
        Magnetic flux dependence of the conductance at the specified energies for the (a) trivial (\(h / h_\mathrm{c} = 0\)) and (b) topological (\(h / h_\mathrm{c} = 2\)) cases, respectively. \(\Phi_0 = h / 2e\), \(t_\mathrm{LW} = 0.1 t_\mathrm{lead}\), \(L_\mathrm{wire} = 1000a\). The inset in (a) is the magnified plot of the same data.
        }
    \label{fig_Tee-Teh-phase}
\end{figure}

Figure~\ref{fig_Tee-Teh-phase} illustrates the AB effect.
In the trivial phase, the conductance shows almost the constant value inside the gap, with the slight Peierls phase dependence of \(2\pi\)-periodicity. In the case of the topological phase, the periodicity is identical to the trivial one. However, the amplitude is more prominent by a few orders of magnitude. These results demonstrate the enhancement of \(2\pi\)-periodic AB effect mediated by topologically protected MZMs mentioned in the previous section. 
The \(2\pi\)-periodicity is a general property of one-dimensional AB interferometers with a grounded superconducting segment, and it is attributed to the processes with sequential Andreev reflection at multiple SN junctions described in Fig.~\ref{fig_setup_process}~(c,~d). These processes both contribute to the \(2\pi\)-periodicity in \(T_\mathrm{ee}\) and \(T_\mathrm{eh}\). More importantly, the contrasting amplitude in the trivial and topological cases results from the absence or presence of MZMs. As explained in Sec.~\ref{section_2piAB}, the zero modes at SN junctions mediate the sequential Andreev scattering processes and amplify the AB oscillation. On the other hand, the absence of low-energy states in the trivial case eliminates the amplification effect.

\subsection{Trivial Bound States}
Now let us consider the effects of low-energy trivial states referred to as ps-ABSs, which appear at SN junctions with smooth gate potentials. In the previous section, we assumed a clean superconductor, where MZMs mediate Andreev reflection and amplify the \(2\pi\)-periodic AB effect. In this case, the absence of MZMs in the trivial phase strongly suppresses the amplitude. Meanwhile, ps-ABSs, a possible cause of the experimentally observed ZBCPs~\cite{2017_Nichele}, can also contribute to Andreev reflection. Here we investigate the effect of ps-ABSs on nonlocal conductance. We compare the topological and trivial bound states in terms of the periodicity, amplitude, and stability of the AB effect.

The trivial ABSs are implemented by introducing smooth gate potentials at the ends of the wire. The spatial variation of the potential and the superconducting gap is
\begin{align}
	V(x) =
    &\frac{V_\mathrm{L}}{2} \left(1 - \tanh(\frac{x - \Delta x_\mathrm{L}}{\delta x_V})\right) \nonumber \\
	&+ \frac{V_\mathrm{R}}{2} \left(1 + \tanh(\frac{x - L_\mathrm{wire} + \Delta x_\mathrm{R}}{\delta x_V})\right), 
    \label{eq_QD_V}
    \\
	\Delta (x) = 
    &\frac{\Delta_0}{2} \left(1 + \tanh(\frac{x - \Delta x_\mathrm{SC}}{\delta x_\Delta})\right) \nonumber \\
	&\times \frac{1}{2} \left(1 - \tanh(\frac{x - L_\mathrm{wire} + \Delta x_\mathrm{SC}}{\delta x_\Delta})\right).
    \label{eq_QD_D}
\end{align}
Parameters in the following calculations are \(\Delta x_\mathrm{L} = \Delta x_\mathrm{R} = 20.5 a\), \(V_\mathrm{L} = V_\mathrm{R} = 0.35 \Delta\), \(\delta x_V = 5a\), \(\delta x_\Delta = 3a\), \(\Delta x_\mathrm{SC} = 17.8a\), \(k_\mathrm{B}T = 0.0015\Delta\), \(t_\mathrm{LW} = 0.25 t_\mathrm{lead}\), \(\mu_\mathrm{wire} = -1.975 t_\mathrm{wire}\), unless stated otherwise.

\begin{figure}[t]
    \includegraphics[width=85mm]{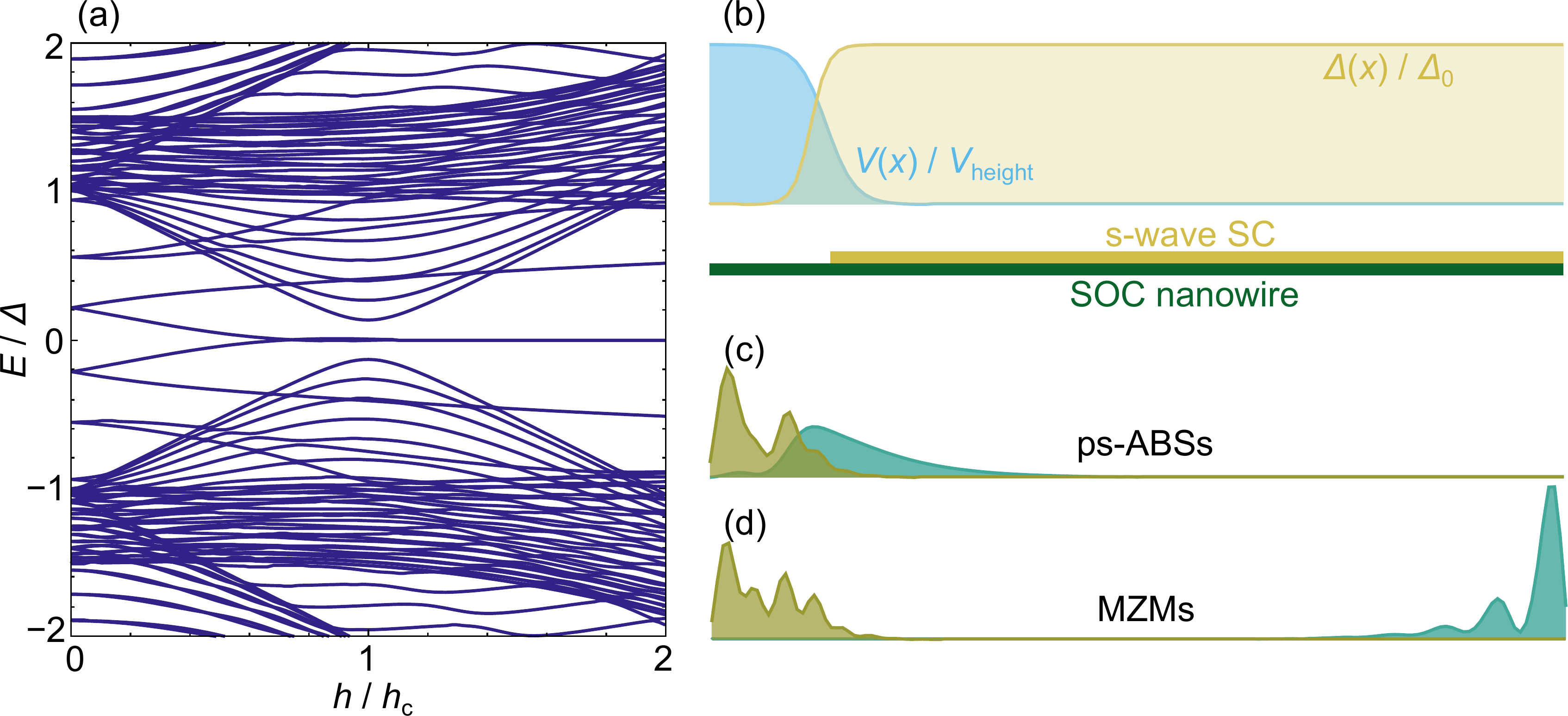}
    \caption{
        (Color online) 
        (a) Magnetic field dependence of the energy eigenvalues of the superconducting nanowire. (b) Spatial variation of the potential \(V(x)\) and the superconducting gap \(\Delta(x)\). (c) and (d) represent the profiles of the squared wavefunctions for (c) ps-ABSs at \(h / h_\mathrm{c} = 0.75\) and (d) MZMs at \(h / h_\mathrm{c} = 1.25\), respectively. \(L_\mathrm{wire} = 151a\).
        }
    \label{fig_psABS-MZM}
\end{figure}

The trivial and topological bound states show contrasting spatial distributions. First, we introduce the gate potential at the left end (see Fig.~\ref{fig_psABS-MZM}~(b)). By diagonalizing the superconducting wire Hamiltonian, we find the eigenfunctions \(\phi_\varepsilon ^{\tau \sigma} (x)\) with eigenenergy \(\varepsilon\), labeled by electron-hole degrees of freedom \(\tau\) and spin \(\sigma\). 
Figure~\ref{fig_psABS-MZM}(c, d) show the distributions of the squared wave functions of ps-ABSs and MZMs in the trivial and topological regime, respectively. The distributions are calculated by 
\begin{equation}
	\abs{\psi_\pm (x)}^2 = \sum_{\tau = \mathrm{e, h}} \sum_{\sigma = \uparrow, \downarrow} \abs{
		\frac{\phi_{\varepsilon}^{\tau \sigma}(x) \pm \phi_{-\varepsilon}^{\tau \sigma}(x)}{\sqrt{2}}
		}^2
    \label{eq_WF}
\end{equation}
These are consistent with the previous research~\cite{2018_Moore_2}.

\begin{figure}[t]
    \includegraphics[width=85mm]{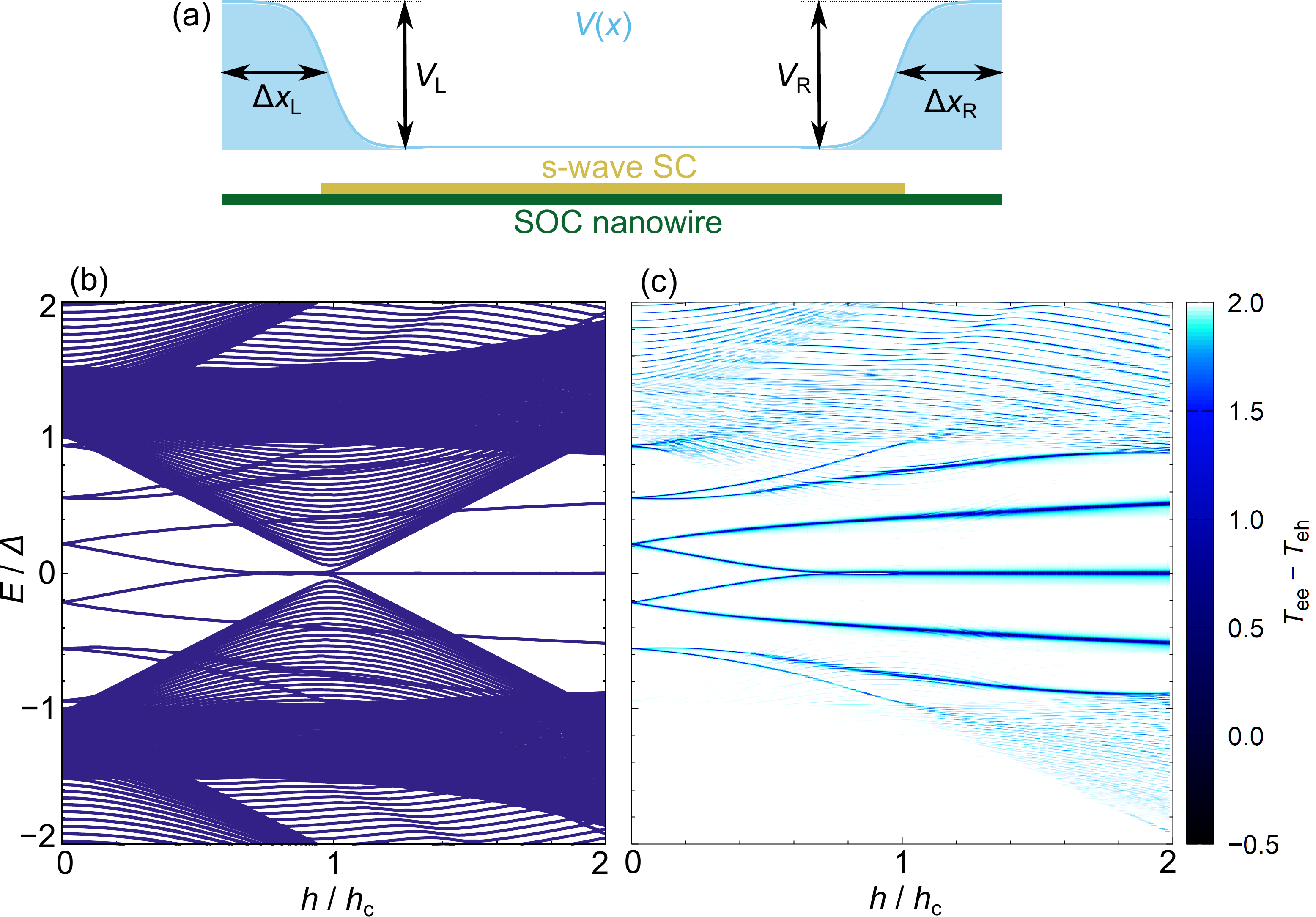}
    \caption{
        (Color online) 
        (a) Spatial variation of the potential. (b) Magnetic field dependence of the energy eigenvalues of the isolated superconductor. (c) Magnetic field dependence of the transmission rate (\(T_\mathrm{ee} - T_\mathrm{eh}\)), or the conductance in the unit of \(e^2 / h\) at zero temperature.
        }
    \label{fig_h-E-G_psABSs}
\end{figure}

\begin{figure}[t]
    \includegraphics[width=85mm]{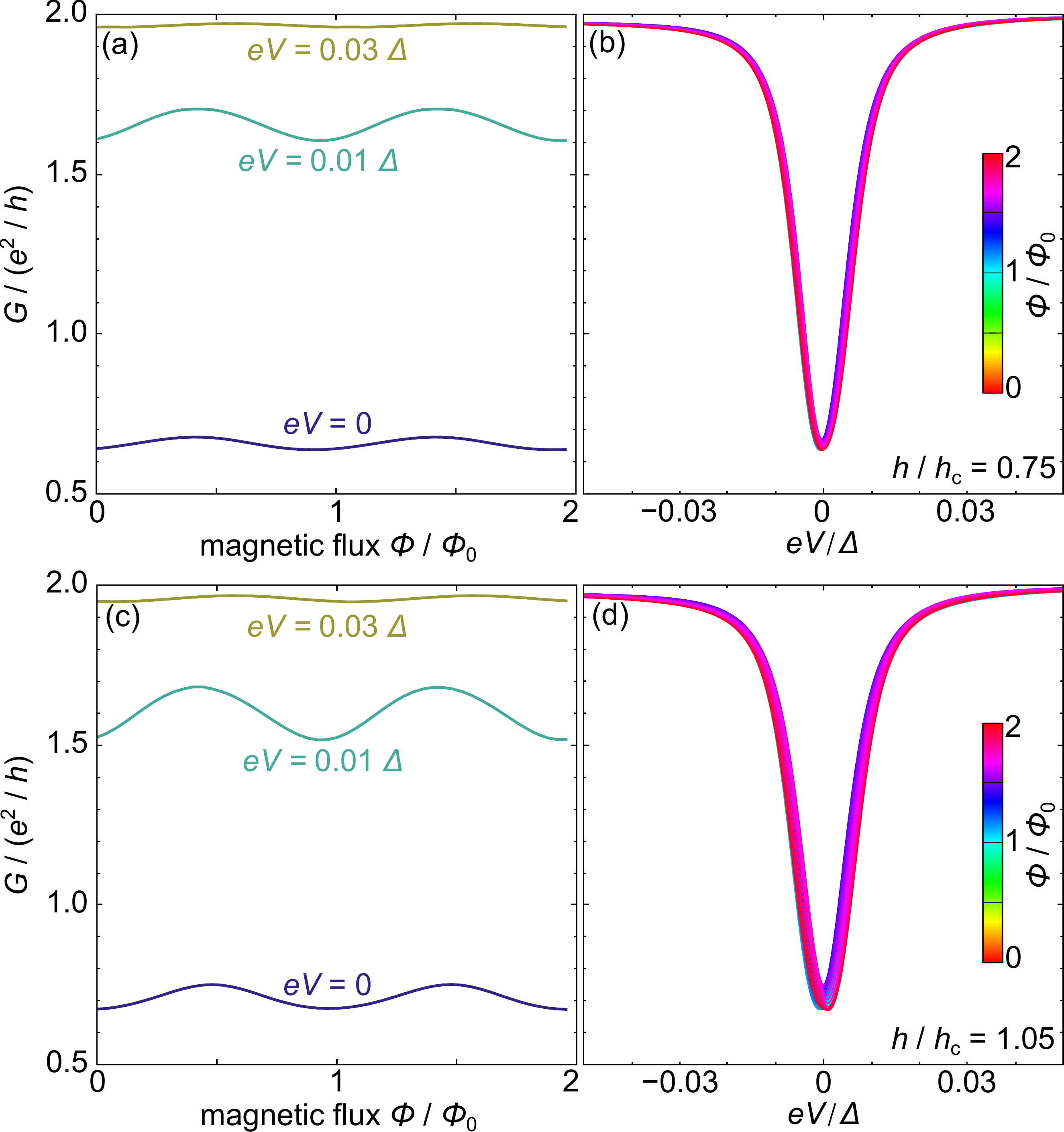}
    \caption{
        (Color online) 
        (a, c) Magnetic flux dependence of the conductance at the specified bias voltage. The voltage dependence is plotted in (b) and (d) with varying the flux (Peierls phase). The top and bottom rows correspond to the trivial (\(h / h_\mathrm{c} = 0.75\)) and topological (\(h / h_\mathrm{c} = 1.05\)) cases, respectively. \(\Phi_0\) is the superconducting magnetic flux quantum \(h / 2e\).
        }
    \label{fig_G-phase_psABSs-MZMs}
\end{figure}

Now we introduce the gate potentials to both ends of the wire and calculate the conductance, as illustrated in Fig.~\ref{fig_h-E-G_psABSs}. The MZMs and the corresponding zero energy structure are observed in the topological regime of \(h > h_\mathrm{c}\). However, unlike the clean case considered in the previous section, the structure extends into the trivial regime, reflecting the presence of ps-ABSs (see Fig.~\ref{fig_h-E-G_psABSs}). 
Figure~\ref{fig_G-phase_psABSs-MZMs} demonstrates the AB effect in each case with topological and trivial bound states. 
The identical periodicity of \(2\pi\) and comparable amplitude in the trivial phase are attributed to the Andreev reflection mediated by ps-ABSs.
These signatures indicate that neither periodicity nor amplitude alone cannot serve as a tool to differentiate trivial and topological bound states. Based on this result, in the following section, we focus on the robustness of the enhanced \(2\pi\)-periodic AB effect in the trivial and topological phase. We show that the parameter dependence of a nonlocal index ``amplitude'', whose definition we introduce later, can distinguish MZMs and ps-ABSs.

\subsection{Robustness}
\begin{figure}[t]
    \includegraphics[width=85mm]{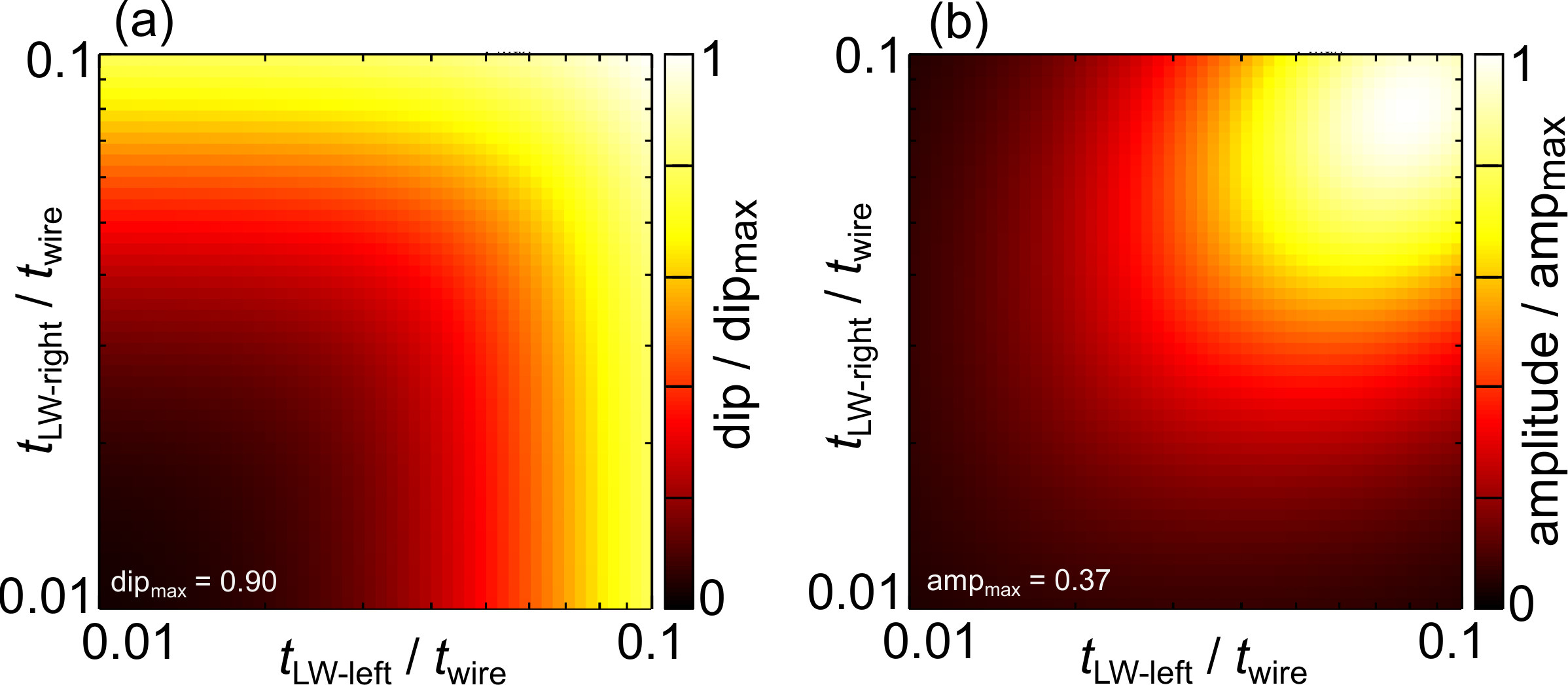}
    \caption{
        (Color online) 
        The two indices, (a) for ``dip'' and (b) for ``amplitude'', are plotted with the varying hoppings at the left and right junctions (\(t_\mathrm{LW\text{-}left}\) and \(t_\mathrm{LW\text{-}right}\), respectively). The superconductor is in the topological phase (\(h / h_\mathrm{c} = 2\)). The superconductor is homogenous (\(\Delta (x)\) is constant and \(V(x) = 0\)). \(\mu_\mathrm{wire} = -2 t_\mathrm{wire}\).
        }
    \label{fig_dip_amp}
\end{figure}

Here we investigate the stability of the \(2\pi\)-periodic AB effect with variations of the potential structure at the ends of the superconducting wire. We consider two indices to characterize the effect: ``dip'' and ``amplitude.'' The former is the depth of valley structures of conductance (\(2 - \bar{G}_\mathrm{ave}\)), and the latter is the width of the \(2\pi\)-periodic oscillation (\(\bar{G}_\mathrm{max} - \bar{G}_\mathrm{min}\)), where
\begin{align}
    \bar{G}_\mathrm{ave} &= \frac{1}{2\Phi_0} \int_0^{2\Phi_0} d \Phi G(eV = 0, \Phi) / (e^2/h), \\
    \bar{G}_\mathrm{max} &= \max_{0 \leq \Phi < 2\Phi_0} G(eV = 0, \Phi) / (e^2/h), \\
    \bar{G}_\mathrm{min} &= \min_{0 \leq \Phi < 2\Phi_0} G(eV = 0, \Phi) / (e^2/h).
\end{align}
Let us discuss how the indices depend on the conditions of the SN junctions.
Figure~\ref{fig_dip_amp} shows contrasting behaviors of ``dip'' and ``amplitude.'' 
The first index ``dip'' remains finite even if either of the SN junctions is switched off, and sequential Andreev reflection at both junctions is unnecessary. Since the index relies on local Andreev reflection at a single SN junction, it corresponds to signatures of local conductance, which only depends on states at either end. Although the index is defined for nonlocal conductance, its value is determined primarily by the local condition. In this sense, the index encodes local information. On the other hand, Fig.~\ref{fig_dip_amp}~(b) shows that ``amplitude'' survives only when both hoppings are finite. This dependence arises because the index relies on sequential Andreev reflection at both junctions. Therefore, ``amplitude'' depends on the condition of the spatially separated ends of the wire, potentially encoding the nonlocal information. In this sense, ``amplitude'' is a nonlocal index.

\begin{figure*}
    \includegraphics[width=\textwidth]{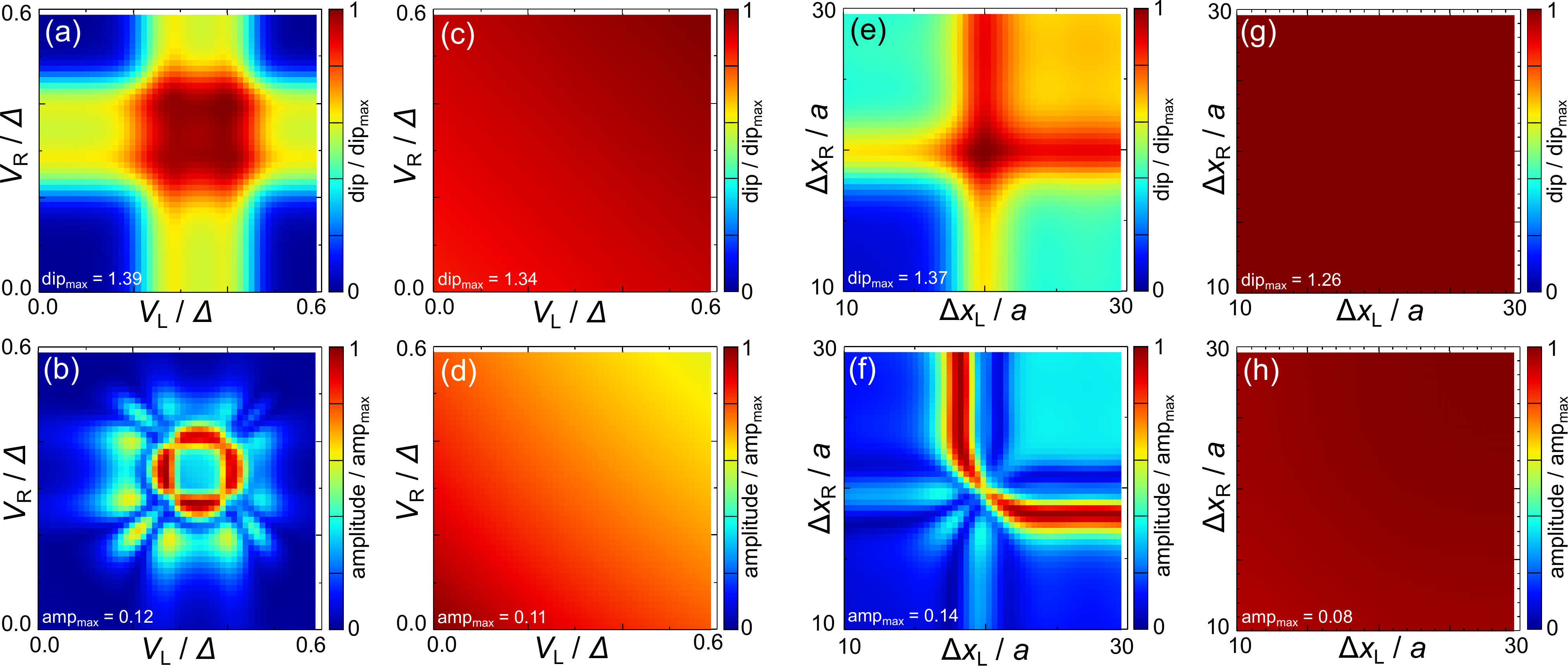}
    \caption{
        (Color online) 
        The top and bottom rows show the stabilities of the indices ``dip'' and ``amplitude,'' respectively, for changes in potential parameters. (a)--(d) demonstrate the robustness against changes in the potential height at the left and right ends. In these cases, the width of the potential profile is \(\Delta x_\mathrm{L} = \Delta x_\mathrm{R} =  20.5a\).
        (e)--(h) show the robustness against changes in the width, with the height set to be \(V_\mathrm{L} = V_\mathrm{R} = 0.35\Delta\).
        The bound states are trivial for (a), (b), (e), (f) with \(h / h_\mathrm{c} = 0.75\) and topological for (c), (d), (g), (h) with \(h / h_\mathrm{c} = 1.25\). The plots are normalized by the maximum values (\(\mathrm{dip_{max}}\) and \(\mathrm{amp_{max}}\)) in each dataset, which are displayed at the bottom left corners of the plots.
    }
    \label{fig_stability}
\end{figure*}

In Fig.~\ref{fig_stability}, we compare the robustness of the two indices for the topological or trivial bound states. 
In the case of MZMs, both indices are robust against variations in potential (see Fig.~\ref{fig_stability}~(c,~d) for ``dip'' and ``amplitude'', respectively), as expected from the topological protection.
The trivial bound states can also exhibit similar behavior in certain parameter regimes for the local index ``dip'' (see Fig.~\ref{fig_stability}~(a)), posing potential difficulty in differentiating ps-ABSs from MZMs. 
These regimes correspond to the case where both junctions independently satisfy the optimal condition for Andreev reflection. The independence suggests an absence of nonlocal information in the index.
On the other hand, ``amplitude'' of ps-ABSs (see Fig.~\ref{fig_stability}~(b)) mimics that of MZMs only in a limited region compared to ``dip.'' Even in the parameter regime where the distinction between trivial and topological bound states by the local index ``dip'' is difficult, ``amplitude'' for ps-ABSs is highly sensitive to variations of the potential profile. This result reflects the nonlocal nature of ``amplitude'', which requires sequential Andreev reflection at both ends of the superconductor.
In the topological case, MZMs are topologically guaranteed to be localized at the ends and remain at zero energy. On the other hand, ps-ABSs are local, and their energies are less stable than those of MZMs. Since sequential Andreev reflection and enhanced \(2\pi\)-periodic AB effect require optimal conditions of multiple SN junctions, the instability of ps-ABSs is more likely to manifest in the nonlocal index than in the local one.

Furthermore, to confirm that our results do not depend on specific model systems, we performed numerical calculations for other system parameters. The results are shown in Appendices~\ref{section_appendix_a} and \ref{section_appendix_b}. As clearly shown in Figs.~\ref{fig_appendix_stability_tlw025} and \ref{fig_appendix_amp-dip_V-symm} in Appendix, the point addressed above holds even for the parameter regime where, as shown in Fig.~\ref{fig_appendix_E_h}, the trivial ps-ABSs appear in the field range much wider than the case discussed in this section. The instability of the trivial bound states in the nonlocal index is confirmed, which implies the generality of the results and conclusions.

\section{Conclusion}
In this paper, we have discussed the \(2\pi\)-periodic AB effect in the context of the Majorana signature in nonlocal transport. This effect is attributed to Andreev reflection at multiple SN junctions and does not require parity restriction. We have demonstrated the enhancement of the effect, where Andreev reflection mediated by MZMs strongly enhances the oscillation amplitude. The processes are robust thanks to the topological protection of MZMs, which guarantees the existence of MZMs at both ends of the superconducting wire.
By introducing the two characteristic indices, we have investigated the influence of trivial bound states. 
The trivial and topological bound states exhibit similar phase dependence in the case with the parameters set ideally for ps-ABSs to mimic MZMs.
However, the robustness against variation of potential profiles demonstrates striking contrast between the trivial and topological bound states. Even in a parameter regime where the local index for ps-ABSs is indistinguishable from the topological one, their nonlocal index, i.e. ``amplitude'', strongly depends on the junction profile, making a sharp contrast to MZMs. The difference can enable a more explicit distinction between the trivial and topological bound states than local conductance measurements. Therefore, the observation of the effect will advance the exploration of their existence.

\begin{acknowledgments}
This work was supported by JST CREST Grant No. JPMJCR19T5, Japan, and the Grant-in-Aid for Scientific Research on Innovative Areas ``Quantum Liquid Crystals (No. JP20H05163 and No. JP22H04480)'' from JSPS of Japan, and JSPS KAKENHI (Grants No. JP20K03860, No. JP20H01857, No. JP21H01039, and No. JP22H01221).
\end{acknowledgments}

\appendix
\section{Numerical Results for Other System Parameters}
\label{section_appendix_a}
\begin{figure}[t]
    \centering
    \includegraphics[width=50mm]{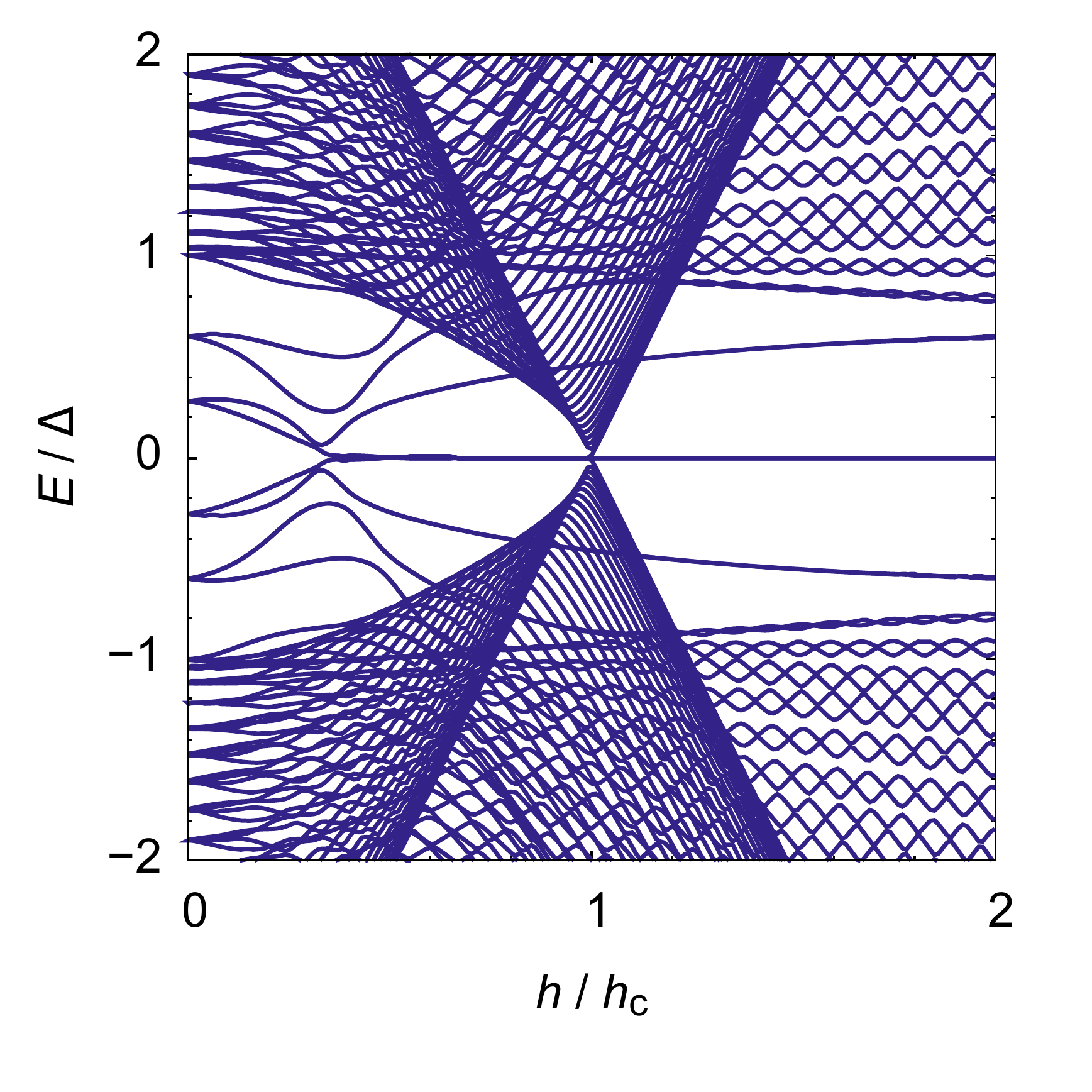}
    \caption{
        (Color online) 
        Magnetic field dependence of the energy eigenvalues of the isolated superconductor. 
        We set \(\mu_\mathrm{wire} = -2t_\mathrm{wire} + 4\Delta\), \(\lambda = 0.182 t_\mathrm{wire}\), \(\Delta = 0.026 t_\mathrm{wire}\), \(\Delta x_\mathrm{L} = \Delta x_\mathrm{R} = 40a\), \(\delta x_V = \delta x_\Delta = 6.5a\), \(\Delta x_\mathrm{SC} = 55a\).
        }
    \label{fig_appendix_E_h}
\end{figure}

Here we show results for other system parameters regimes. We set parameters such that the trivial bound states exist in a wider magnetic field range compared to the case considered in the main text (see Fig.~\ref{fig_appendix_E_h}).
We set the chemical potential \(\mu_\mathrm{wire} = -2t_\mathrm{wire} + 4\Delta\), Rashba spin-orbit coupling \(\lambda = 0.182 t_\mathrm{wire}\), and the induced superconducting gap \(\Delta = 0.026 t_\mathrm{wire}\). 
We set the parameters in \(V(x)\) (Eq.~\eqref{eq_QD_V}) and \(\Delta (x)\) (Eq.~\eqref{eq_QD_D}) as \(\Delta x_\mathrm{L} = \Delta x_\mathrm{R} = 40a\), \(\delta x_V = \delta x_\Delta = 6.5a\),
\(\Delta x_\mathrm{SC} = 55a\).

\begin{figure*}
    \includegraphics[width=\textwidth]{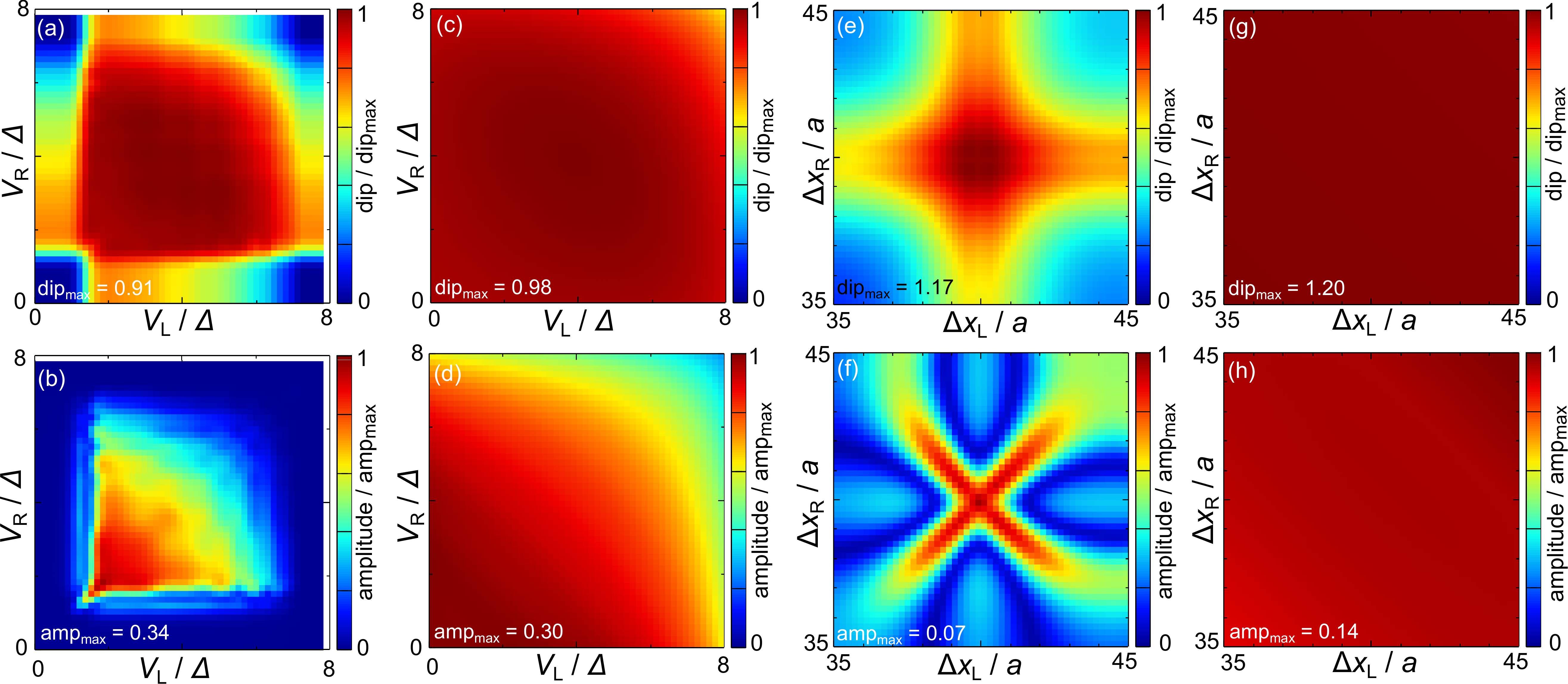}
    \caption{
        (Color online) 
        The top and bottom rows show the color maps of the indices ``dip'' and ``amplitude,'' respectively, as functions of potential parameters. (a)--(d) demonstrate the robustness against changes in the potential height at the left and right ends. In these cases, the width of the potential profile is \(\Delta x_\mathrm{L} = \Delta x_\mathrm{R} = 40a\). \(k_\mathrm{B}T = 0.003 \Delta\).
        (e)--(h) show the robustness against changes in the width, with the height set to be \(V_\mathrm{L} = V_\mathrm{R} = 5.5 \Delta\). \(k_\mathrm{B}T = 0.0015 \Delta\).
        The bound states are trivial for (a), (b), (e), (f) with \(h / h_\mathrm{c} = 0.75\) and topological for (c), (d), (g), (h) with \(h / h_\mathrm{c} = 1.25\). The plots are normalized by the maximum values (\(\mathrm{dip_{max}}\) and \(\mathrm{amp_{max}}\)) in each dataset, which are displayed at the bottom left corners of the plots.
    }
    \label{fig_appendix_stability_tlw025}
\end{figure*}

In Fig.~\ref{fig_appendix_stability_tlw025}, we compare the robustness of the two indices with the topological or trivial bound states. 
As mentioned in the main text, the result shares the same characteristics as Fig.~\ref{fig_stability}: the nonlocal index ``amplitude'' is more sensitive to the variation of the potential profile than the local index. The similarity of the result implies that our conclusions do not rely on the fine-tuning of the system parameters.

\section{Plateau in the Nonlocal Index and Hybridization of MZMs}
\label{section_appendix_b}
\begin{figure*}
    \includegraphics[width=\textwidth]{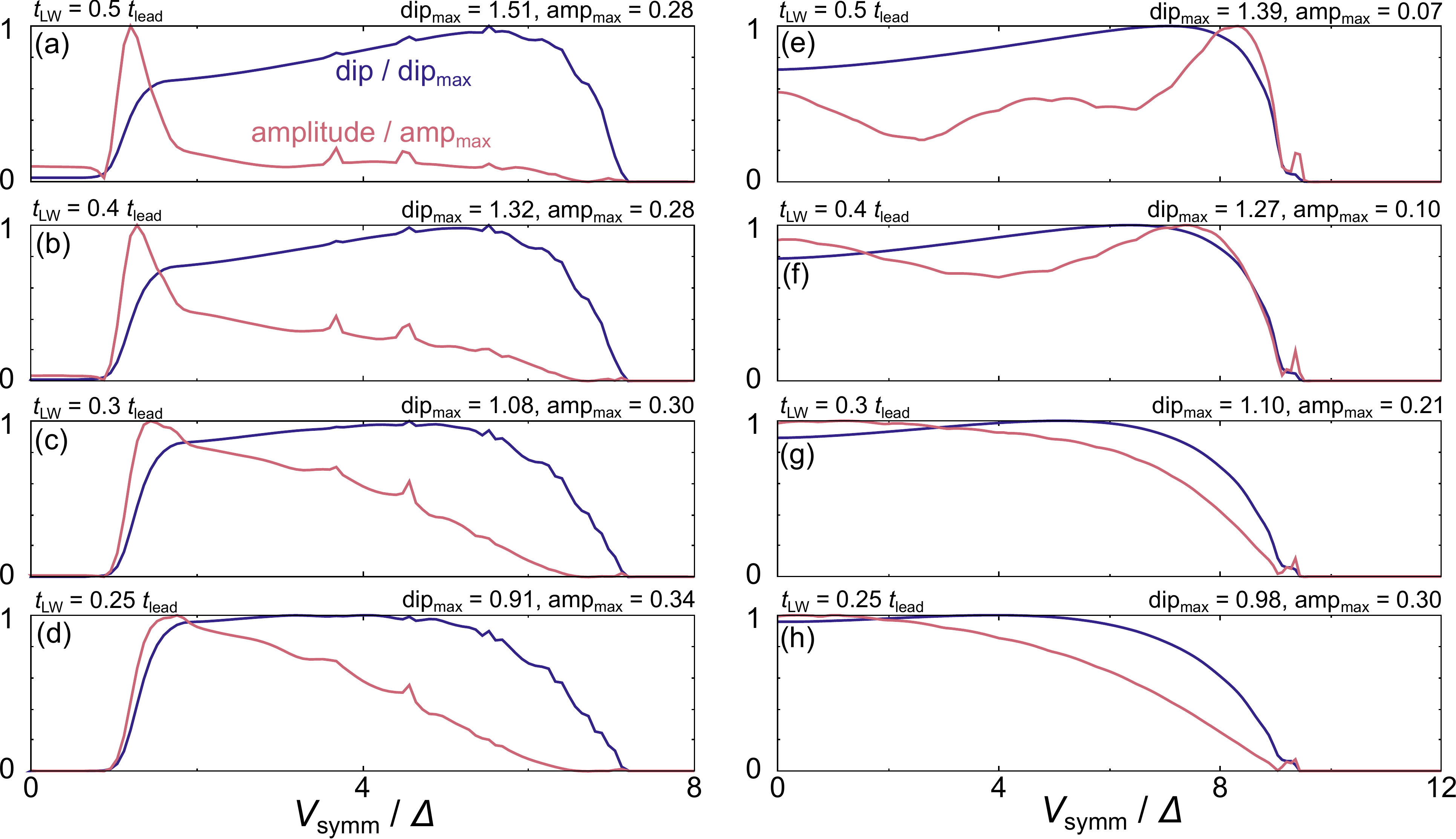}
    \caption{
        (Color online) 
        Potential height dependence of dip and amplitude. The potential heights \(V_\mathrm{L}\) and \(V_\mathrm{R}\) are set to be symmetric, denoted as \(V_\mathrm{symm}\). The vertical axes are normalized by the maximum values in each plot, which are displayed on the top right sides.
        (a)--(d) are for the trivial case with \(h / h_\mathrm{c} = 0.75\) and (e)--(h) are for the topological case with \(h / h_\mathrm{c} = 1.25\).
        The values of the hopping amplitude at the SN junctions are \(t_\mathrm{LW} / t_\mathrm{lead} = 0.5, 0.4, 0.3, 0.25\) for the first to fourth row, respectively.
        \(k_\mathrm{B}T = 0.003 \Delta\).
        }
    \label{fig_appendix_amp-dip_V-symm}
\end{figure*}

In Fig.~\ref{fig_appendix_amp-dip_V-symm}, we show the ``dip'' and ``amplitude'' indices as functions of \(V_\mathrm{symm} \equiv V_\mathrm{L} = V_\mathrm{R}\), with different values of the SN junction coupling \(t_\mathrm{LW}\).
We use the same system parameters as Appendix~\ref{section_appendix_a}. For relatively small values of \(t_\mathrm{LW}\), there exists a parameter regime where both ``dip'' and ``amplitude'' for MZMs are stable (see Fig.~\ref{fig_appendix_amp-dip_V-symm}~(h)). 
For ps-ABSs, the local index ``dip'' appears relatively robust against variation of \(V_\mathrm{L} = V_\mathrm{R} \equiv V_\mathrm{symm}\) (see Fig.~\ref{fig_appendix_amp-dip_V-symm}~(d)). However, the nonlocal index ``amplitude'' is more sensitive to the potential profile, and there is no plateau in ``amplitude'' for ps-ABSs inside the parameter regime where ``dip'' is stable. In fact, the nonlocal index for ps-ABSs decays more rapidly than that of MZMs, presenting a qualitative difference.
In the case of MZMs, the nonlocal index ``amplitude'' is robust against the changes of system parameters provided that \(t_\mathrm{LW}\) is sufficiently small. However, the robustness is spoiled for the larger values of \(t_\mathrm{LW}\) (see Fig.~\ref{fig_appendix_amp-dip_V-symm}~(e)). The probable cause of this fragility is the hybridization of the two MZMs at each end mediated via the metallic lead. The larger the value of \(t_\mathrm{LW}\) is, the stronger the hybridization becomes, leading to the level-splitting. If the energy splitting is larger than the energy scale of the temperature, the AB effect loses the benefit of the topological protection of MZMs, causing the fluctuation. Conversely, the robustness of the nonlocal index can be recovered by adjusting the temperature.

\bibliographystyle{jpsj}
\bibliography{paper_2piAB}

\end{document}